\documentclass{ws-mpla}
\usepackage[super]{cite}
\usepackage{graphicx}
\begin{document}

\markboth{A. Patk\'os}
{Electromagnetic energy transfer in EEMD}

\catchline{}{}{}{}{}

\title{Electromagnetic energy transfer processes\\ in Effective Electro-Magneto Dynamics of axions}

\author{A. PATK\'OS}

\address{Institute of Physics, E\"otv\"os Lor\'and University, P\'azm\'any P\'eter s\'et\'any 1/A\\
Budapest, H-1117, Hungary\\
patkos@galaxy.elte.hu
}

\maketitle

\pub{Received (Day Month Year)}{Revised (Day Month Year)}

\begin{abstract}
Oscillating and dissipative energy exchange between the electromagnetic and axion fields is investigated in an effective electro-magneto dynamical (EEMD) model theory, implying the coexistence of axions with hypothetic magnetic charges. An exact formula is presented for the energy transfer between the electromagnetic and axionic sectors. In a first example we compute analytically the homogeneously oscillating electric and magnetic field configurations generated by the combined action of a constant static magnetic field and a periodically oscillating axion condensate. In the second example the electromagnetic radiative energy loss of a gravitationally bound axion configuration  is computed in the EEMD model.  As a result an asymptotic $\sim t^{1/5}$ temporal increase of the clump size is found also in EEMD. 

\keywords{axion electro-magneto dynamics, effective electric and magnetic charge and current densities, retardation effects}
\end{abstract}

\ccode{PACS Nos.: 11.10Gh,11.10Hi}

\section{Introduction}
Most efforts aiming at the detection of axions, a hypothetical pseudo-Goldstone field providing a most natural theoretical interpretation of the suppression of strong CP-violation, exploit   electromagnetic interactions. Several experiments have searched in the past nearly  40 years for axions covering a wide range of the proposed axion-2-photon coupling parameter space allowing also independent variation of the axion mass\cite{sikivie83,sikivie21}. Negative search results in a large part of the expected axion-photon coupling region push the experiments to achieve orders of magnitude improvements in the sensitivity of the planned new experiments, in order to explore an even weaker regime of the axion-photon interaction.

 Very recently Sokolov and Ringwald \cite{sokolov22,sokolov23} have pointed out that a larger set of axion-photon couplings can be considered if one assumes that the axion field is embedded into an electromagnetic theory extended with magnetic charge densities. For this a variant of the axion model of Kim\cite{kim79} and Shifman, Vainhstein and Zakharov\cite{shifman80} is used where the heavy fermions which couple the axions to gluons and photons are actually dyons. The key observation, which allows a wider operator set in the electromagnetic interactions of axions can be simply stated: such a theory necessarily relies on using two (dual) 4-potentials \cite{schwinger68,zwanziger68} instead of a single one (see also Refs.\refcite{bliokh13,singleton95,govaerts23}). As a result in EEMD  three independent effective axion-photon operators can be introduced instead of a single one. Even if magnetically charged objects (or dyons) are too massive which prevents their observation in our cosmic neighbourhood, still axions represent in EEMD an effective magnetic as well as electric 4-current. Experimental results have to be reinterpreted in this wider framework and specific features would carry distinguishable imprints of magnetically charged particles.  Prospects of observing this "monopole-philing" axion was discussed recently in a review by Ringwald\cite{ringwald23}.

The present note discusses two basic classical electromagnetic effects related to axions as they appear in the wider EEMD framework. First, curiously enough, the effective electric and magnetic currents due to some homogeneous axion configuration and a static uniform magnetic field "induce" time dependent uniform electric and magnetic fields. This effect did not attract much attention in the previous literature. It was observed  in the "maxwellian" version of axion electrodynamics by Masaki {\it et al.}\cite{masaki20}, where it leads merely to a shift of the axion mass. However, as will be argued below, it might become quite dramatic in EEMD, resulting in an exponential contribution to the effective axion potential. From the point of view of the energy transfer this effect leads to a reversible exchange between the axionic and the electromagnetic sectors.  

The other phenomenon to be discussed in this note is the non-resonant energy loss of gravitationally bound axion clumps (axion stars) as described within EEMD. The main question is whether the extended parameter space affects the asymptotic depletion rate of axions.

Dilute axion stars might arise of free particles merely due to their gravitational interaction.  The characteristics of axionic scalar stars has been extensively explored in the literature \cite{tkachev91, kolb93,kolb94,khlopov98,chavanis11a,chavanis11b}. Semi-analytical solutions in the gravity dominated branch were constructed by Eby {\it al.}\cite{eby15}.  Variational technique using the analogy with non-relativistic quantum mechanics has been used to find both the binding energy and the spatial profile of dilute axion stars\cite{guth15}. 

In case of the dilute branch the main mechanism leading to the depletion of the axionic medium is the two-photon decay of axions\cite{sikivie83,sikivie21}.  This effect might be exponentially amplified by parametric resonance \cite{levkov20}. The resonance is extremely narrow, therefore  the momentum spread  of axions typical for gravitationally bound axion clumps has  a great chance to smear it\cite{arza20}.

Strong external magnetic fields are present around neutron stars ($10^4 - 10^{11}$)T, in particular around magnetars ($10^9 - 10^{11}$)T. Non-resonant electromagnetic radiation from axion stars embedded in magnetic fields has been studied intensively \cite{amin21, sen22a}.  A computational algorithm to include the radiation back reaction effect on the axion source into the evolution of dense branch stars was put forward and implemented very recently\cite{sen22}.
The research presented in the second part of this note repeats essentially a recent analysis of the author realised in the framework of the original axion-electromagnetic (maxwellian) theory \cite{patkos23} and applicable in case of dilute stars.

This presentation is organised as follows. In the next section we give a simple and transparent introduction of the classical field theory of EEMD, following the line of thought of Bliokh {\it et al.} \cite{bliokh13}. The extended effective axion-photon interactions will be introduced as part of the corresponding modified set of Maxwell's equations. The conservation of energy (generalised Poynting-theorem) is also formulated there.  In section 3 the exact solution of the modified Maxwell equations in a uniform time-dependent axion condensate and an externally determined constant magnetic field is discussed. Section 4 is devoted to a treatment of electromagnetic radiation from a gravitationally bound axion clump immersed in the homogeneous electromagnetic field configuration derived in the previous section. The time-averaged rate of the electromagnetic energy loss of axion stars is analytically  computed and compared with the conclusions drawn within the original maxwellian theory. Our results are summarized, accompanied with qualitative remarks pointing to some further related phenomena, in the closing section 5.

\section{Axion-photon couplings in effective electro-magneto dynamics}

In axion stars the number density of axions is high, therefore dynamical changes are treated classically. This way one can bystep the consistency considerations of Dirac \cite{dirac31} concerning the quantum embedding of magnetic charges and proceed phenomenologically with dual potentials. First of all, one observes that the relativistic form of Maxwell's equations expressed with a single 4-potential 
\begin{equation}
\partial_\mu F^{\mu\nu}=j^\nu_e,\qquad \partial_\mu F_d^{\mu\nu}=0,\qquad F^{\mu\nu}=\partial^\mu A^\nu-\partial^\nu A^\mu,\qquad F_d^{\mu\nu}=\frac{1}{2}\epsilon^{\mu\nu\rho\sigma}F_{\rho\sigma}
\end{equation}
cannot be completed by placing to the right hand side of the second equation the magnetic current $j^\nu_m$, since it is inconsistent with the Bianchi-identity nature of Faraday's law. This forces the introduction of a second 4-potential, $C^\mu$ fulfilling fully analogous equations:
\begin{equation}
\partial_\mu G^{\mu\nu}=j^\nu_m,\qquad \partial_\mu G_d^{\mu\nu}=0,\qquad G^{\mu\nu}=\partial^\mu C^\nu-\partial^\nu C^\mu,\qquad G_d^{\mu\nu}=\frac{1}{2}\epsilon^{\mu\nu\rho\sigma}G_{\rho\sigma}
\end{equation}
This construction would, however, correspond to two independent vector fields. One complements the dynamical (sourced) equations with a constraint, which forces the two potentials to give rise to the same unique set of field strengths:
\begin{equation}
F_d^{\mu\nu}|_{\bf B,E}=G^{\mu\nu}|_{\bf B,E}.
\label{constraint}
\end{equation}
which in non-relativistic notation leads to the identification (with the metric $g^{\mu\nu}={\textrm {diag}}(1,-1,-1,-1)$)
\begin{equation}
F^{0i}=-E_i,\qquad F^{ij}=-\epsilon^{ijk}B_k,\qquad G^{0i}=-B_i,\qquad G^{ij}=\epsilon^{ijk}E_k.
\label{nonrel-decomposition}
\end{equation}
The dynamical equations follow by independently varying the Lagrange-functional  
\begin{equation}
L_{EMD}=-\frac{1}{8}\left(F_{\mu\nu}F^{\mu\nu}+G_{\mu\nu}G^{\mu\nu}\right)-\frac{1}{2}(j_{e,\mu}A^\mu+j_{m,\mu}C^\mu).
\end{equation}
with respect to both potentials and substituting only in the resulting equations the constraint (\ref{nonrel-decomposition}). The action density of the original theory (defined with a single potential)  is redistributed equally between the two potentials, in order to have the same expression for the energy-momentum density in terms of the field strengths. The coupling of the currents to the potentials is also modified to reproduce the sourced wave equations in the earlier form.

The commonly used extension of this theory by coupling the axion field $a(x)$ to the topological density generated with $A^\mu$ reads\cite{sikivie83,sikivie21} as
\begin{equation}
\Delta L_{aEMD, old}=-\frac{1}{4}g_{aEE}a(x)F_{\mu\nu}F_d^{\mu\nu}.
\end{equation}
By the introduction of the second vector potential one has the freedom to use two further constructs\cite{sokolov22,sokolov23}:
\begin{equation}
\Delta L_{aEMD}=-\frac{1}{8}a(x)\left(g_{aEE}F_{\mu\nu}F_d^{\mu\nu}+g_{aMM}G_{\mu\nu}G_d^{\mu\nu}+2g_{aEM}F_{\mu\nu} G_d^{\mu\nu}\right).
\end{equation}
 Completing Maxwell's equations with the variations of these pieces with respect to $A^\mu$ and $C^\mu$, respectively, one arrives at the following equations:
\begin{eqnarray}
&\displaystyle
\partial_\mu F^{\mu\nu}+g_{aEE}\partial_\mu a(x) F_d^{\mu\nu}+g_{aEM}\partial_\mu a(x)G_d^{\mu\nu}=j^\nu_e\nonumber\\
&\displaystyle
\partial_\mu G^{\mu\nu}+g_{aMM}\partial_\mu a(x) G_d^{\mu\nu}+g_{aEM}\partial_\mu a(x)F_d^{\mu\nu}=j^\nu_m.
\end{eqnarray}
Going over to the non-relativistic decomposition (\ref{nonrel-decomposition}) one writes these equations in the following form:
\begin{eqnarray}
&\displaystyle
\nabla{\bf E}+g_{aEE}{\bf B}\nabla a(x)-g_{aEM}{\bf E}\nabla a(x)=\rho_e,\nonumber\\
&\displaystyle
\nabla{\bf B}-g_{aMM}{\bf E}\nabla a(x)+g_{aEM}{\bf B}\nabla a(x)=\rho_m,\nonumber\\
&\displaystyle
-\dot{\bf E}+\nabla\times{\bf B}-g_{aEE}(\dot a{\bf B}+\nabla a(x)\times{\bf E})+g_{aEM}(\dot a{\bf E}-\nabla a(x)\times {\bf B})={\bf j}_e,\nonumber\\
&\displaystyle
\dot{\bf B}+\nabla\times{\bf E}-g_{aMM}(\dot a{\bf E}-\nabla a(x)\times{\bf B})+g_{aEM}(\dot a{\bf B}+\nabla a(x)\times{\bf E})=-{\bf j}_m.
\label{nonrel-field-equations}
\end{eqnarray}
These equations are consistent with Eqs. (4.18-4.20) of Ref.\refcite{sokolov23}. For later reference we also give the equation of the axion field:
\begin{equation}
\ddot a({\bf x},t)-\nabla^2a({\bf x},t)+m_a^2a({\bf x},t)=(g_{aEE}-g_{aMM}){\bf E}\cdot{\bf B}-g_{aEM}({\bf E}^2-{\bf B}^2).
\label{axion-field-equation}
\end{equation}

The effective axionic charge and current densities can be defined in view of (\ref{nonrel-field-equations}) as 
\begin{eqnarray}
&\displaystyle
\rho_{axion,e}=-g_{aEE}{\bf B}\nabla a(x)+g_{aEM}{\bf E}\nabla a(x),\nonumber\\
&\displaystyle
\rho_{axion,m}=g_{aMM}{\bf E}\nabla a(x)-g_{aEM}{\bf B}\nabla a(x),\nonumber\\
&\displaystyle
{\bf j}_{axion,e}=g_{aEE}(\dot a{\bf B}-\nabla a(x)\times{\bf E})-g_{aEM}(\dot a{\bf E}-\nabla a(x)\times {\bf B})\nonumber\\
&\displaystyle
{\bf j}_{axion,m}=-g_{aMM}(\dot a{\bf E}-\nabla a(x)\times{\bf B})+g_{aEM}(\dot a{\bf B}+\nabla a(x)\times{\bf E}).
\label{effective-charge-current}
\end{eqnarray}
The generalized Poynting-theorem is derived following the standard textbook procedure \cite{jackson98} leading to
\begin{eqnarray}
&\displaystyle
-\int d^3x\left[({\bf j}_{e}+{\bf j}_{axion,e})\cdot {\bf E}(t,{\bf x})+({\bf j}_{m}+{\bf j}_{axion,m})\cdot {\bf B}(t,{\bf x})\right]~~~~~~~~~~~~~~~~~~~~~~~~~~~~~\nonumber\\
&\displaystyle
~~~~~~~~~~~~~~~~~~~~~~~~~~~=\int d{\bf F}\cdot({\bf E}\times{\bf B})+\frac{d}{dt}\int d^3x \frac{1}{2}\left({\bf E}^2(t,{\bf x})+{\bf B}^2(t,{\bf x})\right).
\label{poynting-balance-exact}
\end{eqnarray}
This result is just a slight extension of the relation employed recently in discussions of various phenomena in the original formulation of axion electrodynamics \cite{patkos22,paixao22}.  On the right hand side one recognizes the familiar expressions of the rate of change of the electromagnetic energy and the power of electromagnetic radiation.\footnote{Alternative fazor formulation of the Poynting theorem in terms of complex fields allows in the complex valued energy balance equation the separation of the reactive and dissipative parts of the energy transfer. \cite{tobar22,tobar23}}  Therefore it is natural to identify the left hand side with the negative rate of change of the axion energy.

 Indeed, substituting into the left hand side the axionic currents from (\ref{effective-charge-current}) one finds
\begin{eqnarray}
&\displaystyle
-\int d^3x\left[{\bf j}_{axion,e}\cdot {\bf E}(t,{\bf x})+{\bf j}_{axion,m}\cdot {\bf B}(t,{\bf x})\right]\nonumber\\
&\displaystyle
=-\int d^3x\dot a\left[(g_{aEE}-g_{aMM}){\bf E}\cdot{\bf B}-g_{aEM}({\bf E}^2-{\bf B}^2)\right],
\end{eqnarray}
which by (\ref{axion-field-equation}) equals the rate of change of the axionic energy content
\begin{equation}
-\frac{d}{dt}\int d^3x\frac{1}{2}\left(\dot a^2+(\nabla a)^2+m_a^2a^2\right).
\end{equation}

\section{Time-dependent electromagnetic fields induced by a homogeneous axion condensate and a constant magnetic field}

In this section we consider the emergence of spatially uniform time-dependent electric and magnetic fields under the combined action of an externally determined space-time independent magnetic field ${\bf B}_0$ and an arbitrary (though periodically) varying homogeneous axion $a_0(t)$ field. The time dependence of $a_0(t)$ is determined next consistently taking into account the full electromagnetic field in its equation of motion. The special case of a small amplitude oscillatory motion of the axion will be analyzed in more detail.

The field equations (\ref{nonrel-field-equations}) for the emerging field strengths ${\bf B}(t), {\bf E}(t)$ simplify to the following form
\begin{eqnarray}
&\displaystyle
\dot{\bf E}(t)=-g_{aEE}\dot a_0(t)({\bf B}_0+{\bf B}(t))+g_{aEM}\dot a_0(t){\bf E}(t)\nonumber\\
&\displaystyle
\dot{\bf B}(t)=g_{aMM}\dot a_0(t){\bf E}(t)-g_{aEM}\dot a_0(t)({\bf B}_0+{\bf B}(t)).
\label{homogeneous-field-eq}
\end{eqnarray}
The solutions can be parametrized with help of the single available vector ${\bf B}_0$ as
\begin{equation}
{\bf E}=f_E(a_0(t)){\bf B}_0,\qquad {\bf B}=f_B(a_0(t)){\bf B}_0
\end{equation}
supplemented with the initial conditions
\begin{equation}
f_E(a_0(t_0)=a_i)=0,\qquad f_B(a_0(t_0)=a_i)=0.
\label{initial-condition}
\end{equation}
The initial value $a_i$ is left at this stage undetermined.  After substituting this into (\ref{homogeneous-field-eq}) one finds the equations
\begin{eqnarray}
&\displaystyle
f^\prime_E(a_0)+g_{aEE}(1+f_B(a_0))-g_{aEM}f_E(a_0)=0, \nonumber\\
&\displaystyle
 f^\prime_B(a_0)-g_{aMM}f_E(a_0)+g_{aEM}(1+f_B(a_0))=0,
\end{eqnarray}
where $f^\prime(a_0)=df(a_0)/da_0$.  This set obviously has two solutions exponentially depending on $a_0(t)$, 
\begin{equation}
f_\cdot=f_{\cdot 0}e^{\lambda a_0},\qquad \lambda^2=\lambda_\pm^2=g^2_{aEM}-g_{aEE}g_{aMM}.
\end{equation}
We assume below that the sign of $\lambda^2$ is positive, although one can repeat the procedure with imaginary roots without any complication. 
The solution fulfilling the initial conditions (\ref{initial-condition}) has the following explicit expression:
\begin{eqnarray}
&\displaystyle
f_E=-\frac{g_{aEE}}{\lambda}\sinh [\lambda( a_0(t)-a_i)],\nonumber\\
&\displaystyle
f_B=\cosh[\lambda(a_0(t)-a_i)]-1-\frac{g_{aEM}}{\lambda}\sinh[\lambda(a_0(t)-a_i)].
\end{eqnarray}
It is noteworthy that in the "maxwellian" axion electromagnetism, where only $g_{aEE}\neq 0$ (and therefore $\lambda\rightarrow 0$), the exact solution is given with a purely linear $a_0$-dependence of $F_E(a_0)$. 

Having in mind that the above solution for the electric and magnetic field strength parametrically depends on the homogeneous axion field, one can find the effective equation which determines its time dependence by evaluating the right hand side of (\ref{axion-field-equation}). The resulting expression can be interpreted as the $a_0$-derivative of an effective contribution $U_a(a_0)$ to the effective potential $V_{axion}$ of the axion. 
 The parameter $a_i$ is found as the static equilibrium point of the resulting $V_{axion}(a_0)=m_a^2a_0^2+U_a(a_0)$.
The "induced" potential $U_a(a_0)$ modifies the simple harmonic time dependence of the usually assumed coherent motion of free axions $a_0(t)=A\sin(m_at)$. The underlying variation of the electromagnetic fields on the other hand produces periodic fluctuations in the initially constant magnetic field. All this could be observable in the motion through such a medium of electrically charged objects, magnetic dipoles, etc.

In order we could proceed further analytically, we restrict the investigation to the harmonic case by assuming small amplitude oscillations:
\begin{equation}
|\lambda(a_0(t)-a_i)|<<1.
\end{equation}
Expanding the right hand side of (\ref{axion-field-equation}) up to linear terms one finds
\begin{equation}
{\bf B}(t)=-g_{aEM}\delta a(t){\bf B}_0,\qquad {\bf E}(t)= -g_{aEE}\delta a(t){\bf B}_0
\label{time-dependent-space-independent-fields}
\end{equation}
 with the following equation for $\delta a(t)=a_0(t)-a_i$:
\begin{equation}
\delta\ddot a+m^2_a\delta a=-[(g_{aEE}-g_{aMM})g_{aEE}+2g_{aEM}^2]B_0^2\delta a,\qquad
a_i= g_{aEM}\frac{B_0^2}{m_a^2}.
\end{equation}
The effect of the electromagnetic background for such small amplitude coherent oscillations consists in a shift of the eigenfreqency to
\begin{equation}
M_a^2=m_a^2+[(g_{aEE}-g_{aMM})g_{aEE}+2g_{aEM}^2]B_0^2.
\end{equation}
This expression generalizes the effect found by Masaki {\it et al.} \cite{masaki20} in the maxwellian version of axion electromagnetism. It represents a selfsustaining configuration with electric and magnetic field strengths parallel to the externally imposed magnetic field. Consequently it does not release electromagnetic radiation. 

 In the next section a finite size clump is constructed out of the above coherently oscillating configuration stabilized by Newtonian gravitational interaction among the constituent axions. This object does  not represent any stationary solution of the coupled axionic Maxwell-equations. Electromagnetic radiation will emerge due to the oscillating source densities derived with the profile function of the clump.

\section{EEMD theory of electromagnetic radiation of axions in static magnetic field}

Irreversible energy dissipation from finite density axion configurations occurs in form of electromagnetic radiation\cite{sikivie83,levkov20}. For its characterisation in EEMD one relies on an effective theory with axions coherently moving in a gravitationally bound axion clump, also called {\it dilute axion star}. The effective potential of the theory which includes the effect of the background electromagnetic fields reads as
\begin{eqnarray}
&\displaystyle
U_{eff}=\frac{1}{2}\left(\dot a^2+(\nabla a)^2+M_a^2a^2\right)+U_{grav}\nonumber\\
&\displaystyle
U_{grav}=-\frac{G_N}{2}\int d^3x\int d^3y\frac{\rho_{axion}({\bf x})\rho_{axion}({\bf y})}{|x-y|}.
\end{eqnarray}.
One can search for the solution in the form
\begin{equation}
a({\bf x},t)=\frac{1}{\sqrt{2M_a}}\left(e^{-iM_at}\psi({\bf x},t)+e^{iM_at}\psi^*({\bf x},t)\right),
\qquad \psi({\bf x},t)=e^{-i\mu_gt}\phi({\bf x}).
\end{equation}
Assuming slow time-variation of $\psi({\bf x},t)$ one finds a Schrodinger-like eigenvalue equation for the gravitational binding energy $\mu_g$. The eigenfunction $\phi({\bf x})$ is normalized on the number $N$ of the axions in the clump, which allows the indentification\footnote{The uniform constant energy density $m_a^2a_i^2/2$ does not enter the gravitational energy, see Olbers-paradox!}  $\rho_{axion}=(M_a+\mu_g)|\phi({\bf x})|^2$.  Then
one can repeat the variational characterisation  of the gravitational binding energy proposed in Ref.\refcite{guth15} (succintly summarised in Ref.\refcite{patkos23}) with the trial function
\begin{equation}
\phi({\bf x})=wF(\xi),\qquad \xi=\frac{|{\bf x}|}{R},\qquad w^2=\frac{N}{C_2R^3}, \qquad 
C_2=\int \frac{d^3x}{R^3} F^2(\xi).
\end{equation}
$F(\xi)$ is called the profile function and in its argument $R$ is a length controlling the radial extension of the axion star. 
The estimate of the "ground-state energy" after minimizing with respect to $R$ at fixed axion number $N$ is proportional to $\sim M_a(G_NM_a^2N)^2$ with a coefficient of $\le {\cal O}(10^2)$, depending on the actual geometry of the profile. It was estimated in Ref.\refcite{guth15} and is much smaller than $M_a$ for  $N\sim 10^{61}$ even with $M_a\sim 10^{-13}$GeV. In this case the specific gravitational binding energy is negligible relative to the effective axion mass and the full mass of the clump is well approximated by $NM_a$. The expression of the axion field to be used below looks like
\begin{equation}
a({\bf x},t)=\sqrt{\frac{2}{M_a}}wF(\xi)\cos (M_at).
\end{equation}

 The electromagnetic radiation emission from the clump to be described below damps the amplitude of the axion oscillations. This is a slow process, which means that $N(t)$ will slowly decrease. In a sort of adiabatic approximation one assumes that every moment the profile of the clump will be the same with the actual axion number. The rate of decrease reads off the energy balance between the axion and the electromagnetic sector.

In the electromagnetic field strengths in addition to ${\bf E}(t), {\bf B}(t)$ determined in the previous section only the radiation fields ${\bf e}(t,{\bf x}), {\bf b}(t,{\bf x})$ are taken into account beyound the static homogeneous field ${\bf B}_0$: 
\begin{equation}
{\bf B}_{full}(t,{\bf x})={\bf B}(t)+{\bf B}_0+{\bf b}(t,{\bf x}),\quad {\bf E}_{full}(t,{\bf x})={\bf E}(t)+{\bf e}(t,{\bf x}).
\end{equation}
In the effective axion currents we use explicitly the contribution from the background fields (\ref{time-dependent-space-independent-fields}) induced by the constant magnetic field 
\begin{eqnarray}
&\displaystyle
\rho_{axion,e}=-g_{aEE}{\bf B}_0\nabla a(x),\qquad \rho_{axion,m}=[-g_{aEM}+\lambda^2a]{\bf B}_0\nabla a,\nonumber\\
&\displaystyle
{\bf j}_{axion,e}=g_{aEE}\dot a{\bf B}_0+[g_{aEM}+(g_{aEE}^2-g_{aEM}^2)a]\nabla a\times{\bf B}_0,\nonumber\\
&\displaystyle
{\bf j}_{axion,m}=[g_{aMM}-g_{aEM}(g_{aMM}+g_{aEE})a]\nabla a(x)\times{\bf B}_0+[g_{aEM}-\lambda^2a]\dot a{\bf B}_0.
\label{effective-current-densities}
\end{eqnarray}
Notice that the continuity equations expressing separate conservation of the magnetic and electric effective charges are fulfilled away from the external current ${\bf J}_0$ generating ${\bf B}_0$, where $\nabla a(x)\times{\bf B}_0=\nabla\times (a{\bf B}_0)$. The field equations of the vector potentials which determine the radiation fields in the Lorentz gauge ($\partial_\mu A^\mu=\partial_\mu C^\mu=0$) become inhomogeneous wave-equations, with sources given in (\ref{effective-current-densities}) :
\begin{equation}
\square A^\mu=j_{axion,e}^\mu,\qquad \square C^\mu=j_{axion,m}^\mu, 
\end{equation}
 We use below their retarded solutions. Here we note that these solutions are periodic in time due to the harmonic axion motion in the source density. Terms quadratic in the axion field generate the second harmonics in the radiation field.

The following discussion will concentrate on the energy power balance :
\begin{eqnarray}
&\displaystyle
-\int d^3x\left({\bf j}_{e,eff}\cdot {\bf e}(t,{\bf x})+{\bf j}_{m,eff}\cdot {\bf b}(t,{\bf x})\right)=-\frac{dE_{axion}}{dt}~~~~~~~~~~~~~~~~~~~~~~~~~~~~~\nonumber\\
&\displaystyle
~~~~~~~~~~~~~~~~~~~~~~~~~~~=\int d{\bf F}\cdot({\bf e}\times{\bf b})+\frac{d}{dt}\int d^3x \frac{1}{2}\left({\bf e}^2(t,{\bf x})+{\bf b}^2(t,{\bf x})\right).
\label{poynting-balance}
\end{eqnarray}
We are interested in the non-vanishing time average of the first line since it determines the rate of energy loss of the axion sector. 
Using
\begin{equation}
{\bf e}=-\dot{\bf A}-\nabla A^0,\qquad {\bf b}=-\dot{\bf C}-\nabla C^0
\end{equation}
with help of the retarded solutions of $A^\mu, C^\mu$ one finds an expression for the rate on the left hand side of (\ref{poynting-balance}). For the sake of simpler expressions starting from here we omit the radiation in the second harmonics, although there is no conceptual obstacle  incomputing also the pwer emitted in this mode. With this restriction theelectric and magnetic power becomes a quadratic functional of the axion field. A great advantage for the present article that  one can rely on the same operations with the potentials as was done in Ref.\refcite{patkos23} in the maxwellian axion electrodynamics. and arrive at
\begin{eqnarray}
&\displaystyle
\int d^3x\left({\bf j}_{e,eff}\cdot {\bf e}(t,{\bf x})+{\bf j}_{m,eff}\cdot {\bf b}(t,{\bf x})\right)~~~~~~~~~~~~~~~~~~~~~~~~~~~~~~~\nonumber\\
&\displaystyle
=-\int d^3x\int d^3x^\prime\frac{1}{|{\bf x}-{\bf x}^\prime|}\Biggl[{\bf j}_{e,eff}({\bf x},t)\cdot\frac{\partial}{\partial t}{\bf j}_{e,eff}({\bf x}^\prime,t-|{\bf x}-{\bf x}^\prime|)\nonumber\\
&\displaystyle
~~~~~~~~~~~~~~~~~~~~~~~~~~~~~~~~~~~~~~~~~~~~+
\frac{\partial}{\partial t}\rho_{e,eff}({\bf x},t)\rho_{e,eff}({\bf x}^\prime,t-|{\bf x}-{\bf x}^\prime|)\Biggr]\nonumber\\
&\displaystyle
-\int d^3x\int d^3x^\prime\frac{1}{|{\bf x}-{\bf x}^\prime|}\Biggl[{\bf j}_{m,eff}({\bf x},t)\cdot\frac{\partial}{\partial t}{\bf j}_{m,eff}({\bf x}^\prime,t-|{\bf x}-{\bf x}^\prime|)\nonumber\\
&\displaystyle
~~~~~~~~~~~~~~~~~~~~~~~~~~~~~~~~~~~~~~~~~~~~+
\frac{\partial}{\partial t}\rho_{m,eff}({\bf x},t)\rho_{m,eff}({\bf x}^\prime,t-|{\bf x}-{\bf x}^\prime|)\Biggr].
\end{eqnarray}

Time averages (over the oscillation period $T=2\pi/M_a$) appearing in the axion's radiation power after substituting this parametrisation into the expressions of the effective charge and current densities (\ref{effective-current-densities}) make use of the following three combinations:
\begin{eqnarray}
&\displaystyle
\overline{\dot a({\bf x},t)\ddot a({\bf x}^\prime,t-|{\bf x}-{\bf x}^\prime|)}^T
=M_a^2w^2F(\xi)F(\xi^\prime)\sin(M_a|{\bf x}-{\bf x}^\prime|),\nonumber\\
&\displaystyle
\overline{\partial_ia({\bf x},t)\partial_j\dot a({\bf x}^\prime,t-|{\bf x}-{\bf x}^\prime|)}^T=-\overline{\partial_i\dot a({\bf x},t)\partial_ja({\bf x}^\prime,t-|{\bf x}-{\bf x}^\prime|)}^T\nonumber\\
&\displaystyle
=\frac{w^2}{R^2}F^\prime(\xi)F^\prime(\xi^\prime)\sin(M_a|{\bf x}-{\bf x}^\prime|)\hat x_i\hat x^\prime_j.
\end{eqnarray}
The explicit expressions for the power of the electric and magnetic contributions have a quite compact form:
\begin{eqnarray}
&\displaystyle
\overline{\int d^3x{\bf j}_{ax}^e{\bf e}}^T=-\int d^3x\int d^3x^\prime\frac{w^2B_0^2}{|{\bf x}-{\bf x}^\prime|}\sin(M_a|{\bf x}-{\bf x}^\prime|)\Bigl[
g_{aEE}^2M_a^2F(\xi)F(\xi^\prime)\nonumber\\
&\displaystyle
-F^\prime(\xi)F^\prime(\xi^\prime)
\frac{1}{R^2}\left(g_{aEE}^2(\hat{\bf x}{\bf n}_B)(\hat{\bf x}^\prime{\bf n}_B)-g_{aEM}^2\hat x_i\hat x_j^\prime(\delta_{ij}-n_{Bi}n_{Bj})\right) \Bigr],\nonumber\\
&\displaystyle
\overline{\int d^3x{\bf j}_{ax}^m{\bf b}}^T=-\int d^3x\int d^3x^\prime\frac{w^2B_0^2}{|{\bf x}-{\bf x}^\prime|}\sin(M_a|{\bf x}-{\bf x}^\prime|)\Bigl[
g_{aEM}^2M_a^2F(\xi)F(\xi^\prime)\nonumber\\
&\displaystyle
-F^\prime(\xi)F^\prime(\xi^\prime)
\frac{1}{R^2}\left(g_{aEM}^2(\hat{\bf x}{\bf n}_B)(\hat{\bf x}^\prime{\bf n}_B)-g_{aMM}^2\hat x_i\hat x_j^\prime(\delta_{ij}-n_{Bi}n_{Bj})\right) \Bigr].
\end{eqnarray}
Here ${\bf n}_B$ denotes the unit vector pointing in the direction of ${\bf B}_0$, and $\hat {\bf x}={\bf x}/|{\bf x}|$.
Using the decomposition of the $|{\bf x}-{\bf x}^\prime|$ dependent pieces of the integrands in terms of spherical harmonics as has been exploited in Ref.\refcite{patkos23}
one can perform the angular integrations and arrive at the following expression for the electromagnetic energy loss of the axion clump with unspecified axion profile function:
\begin{equation}
\overline{\frac{dE_{ax}}{dt}}^T=-\frac{NB_0^2X^3}{C_2}\left[(g_{aEE}^2+g_{aEM}^2)I_{mag}^2-\frac{1}{3X^2}(g_{aEE}^2-g_{aEM}^2-2g_{aMM}^2)I_{el}^2\right],
\end{equation}
where  $X=M_aR$ characterizes the size of the clump in units of $M_a^{-1}$. The quantities $I_{mag}, I_{el}$ depending on the profile of the axion star are the same as in case of the single axion-photon coupling treated in Ref.\refcite{patkos23}:
\begin{equation}
I_{mag}=4\pi\int d\xi \xi^2\frac{\sin(X\xi)}{X\xi}F(\xi),\qquad I_{el}=4\pi\int d\xi\xi^2\left(\frac{\sin(X\xi)}{(X\xi)^2}-\frac{\cos(X\xi)}{X\xi}\right)F^\prime(\xi).
\end{equation}
It is somewhat unexpected that the new couplings formally might turn negative the coefficient of $I_{el}^2$ which is clearly forbidden physically.  This condition imposes a constraint on the relative magnitude of the couplings. However, if one accepts the hierarchy,
$g_{aMM}^2>>g_{aEM}^2>>g_{aEE}^2$, dictated by the relative size of the respective Witten-effects\cite{sokolov23}, clearly both contributions deplenish the axion content of the clump. 

The $X$-dependence of the right hand side  for asymptotically large $X$ values relevant for dilute axion stars  is easily found when one introduces the rescaled variable $u=X\xi$ for the integrals $I_{mag}, I_{el}$. 
In Ref.\refcite{patkos23} it was shown that $I_{mag}\sim X^{-4}$. Assuming finite extension of the clump $\xi\in(0,\Lambda)$ and the boundary conditions $F(\Lambda)=F^\prime(\Lambda)=0$ for the profile, one partial integration leads to an exact relation between $I_{mag}$ and $I_{el}$:
\begin{equation}
I_{mag}=\frac{4\pi}{X^4}\int_0^{X\Lambda}(\sin u-u\cos u)F^\prime(u/X)=-\frac{1}{X}I_{el}\sim X^{-4},\qquad X\rightarrow\infty.
\end{equation}
The reinterpretation of the average energy loss power $\overline{dE_a/dt}^T$ as the rate of depletion $M_a\overline{dN(t)/dt}^T$ of the axion number in the clump leads to the same $N_{axion}\sim t^{-1/5}$ rule as found before introducing the new couplings.

\section{Summary and outlook}

The first phenomenon discussed in this paper discloses the rather dynamical nature of the medium consisting of combined homogeneous electromagnetic and axion fields in EEMD (section 3). It could affect the synchrontron radiation arising from electric charges orbiting around neutron stars, if an axion condensate is also present. In the second part (section 4) we extended the treatment of non-resonant electromagnetic radiation from dilute axion stars to the linearized theory of effective electro-magneto dynamics of axions. The analytic computation of the radiation power led to the same asymptotic result for the rate of change of the axion number of a clump due to non-resonant electromagnetic emission as was found previously in the original version of axion electrodynamics\cite{patkos23}. Along the same lines also the emission power in higher harmonics could be computed in case of any phenomenological interest. The stability of wave propagation through an axion condensate allowing both resonant axion decay and axion-photon conversion in presence of static magnetic field\cite{masaki20}  represents a greater challenge in the EEMD framework and is left for future investigation.

\section*{Acknowledgement}
I thank Anton Sokolov for a clarifiying correspondence on current conservation in EEMD.

\end{document}